\documentclass[aps,prb,preprint,twocolumn,10pt,tightenlines]{revtex4}
\usepackage{amsmath}    
\usepackage{graphicx}   
\usepackage{rotating}
\usepackage{overpic}
\usepackage{bm}
\usepackage{upgreek}
\usepackage[caption=false]{subfig}

\usepackage[utf8]{inputenc}

\newcommand{\eg}{\textit{e.g.}~}
\newcommand{\ie}{\textit{i.e.}~}
\begin{document}

\title{Mechanical approach to surface tension and capillary phenomena}
\author{Marc Durand}
  \affiliation{Mati\`{e}re et Syst\`{e}mes Complexes (MSC), UMR 7057 CNRS \& Universit\'{e}
de Paris, 10 rue Alice Domon et L\'{e}onie Duquet, 75205 Paris Cedex 13, France, EU}
 \email{marc.durand@univ-paris-diderot.fr}   
\date{\today}

\begin{abstract}
Many textbooks dealing with surface tension favor the thermodynamic approach (minimization of some thermodynamic potential such as free energy) over the mechanical approach (balance of forces) to describe capillary phenomena, stating that the latter is flawed and misleading.  
Yet,  mechanical approach is more intuitive for students than free energy minimization, and does not require any knowledge of thermodynamics. In this paper we show that capillary phenomena can be unmistakably described using the mechanical approach, as long as the system on which the forces act is
properly defined. 
After reminding the microscopic origin of a tangential tensile force at the interface, we derive the Young-Dupré equation, emphasizing that this relation should be interpreted as an interface condition at the contact line, rather than a force balance equation. This correct interpretation avoids misidentification of capillary forces acting on a given system. Moreover, we show that a reliable method to correctly identify the acting forces is to define a control volume that does not embed any contact line on its surface.
Finally, as an illustration of this method, we apply the mechanical approach in a variety of ways on a classic example: the derivation of the equilibrium height of capillary rise (Jurin's law).
\end{abstract}
\maketitle


Surface tension and capillary phenomena encompass phenomena in which the interface between to immiscible fluids acts as a thin elastic sheet \cite{de_Gennes}.
This is a standard topic taught at the undergraduate level. Despite its apparent simplicity and its many manifestations in everyday life, this subject is not exempt from traps and misunderstandings \cite{Berry, Marchand}.

In particular, there are two standard issues that students (and teachers) must face
when addressing this topic: \textit{i)} how to explain the tangential orientation of surface tension force from a microscopic perspective ? \textit{ii)} How to properly identify the capillary forces acting on a given system ?
%
%
In most textbooks these points are eluded, and the thermodynamic approach (minimization of suited thermodynamic potential) is favored over the mechanical approach (balance of forces) to describe capillary phenomena. In some of them students are even warned against mechanical approach, questioning the interpretation of surface tension as a force per unit length \cite{Champion, Adam}. However, without using the concept of a tensile
force parallel to the surface, it is very difficult to explain how objects which are significantly more dense than water (\eg a paperclip or a pin) can ``float'' on the water surface \cite{Vella_2005}.


In this paper we show that capillary phenomena can be undoubtedly described using mechanical approach. Indeed, mistakes do not result from the mechanical approach \textit{per se}, but rather from the ambiguity as to the delimitation of the system under study, and thus in the identification of the forces exerted on it. 
%
%
After reminding the microscopic origin of the tangential orientation of surface tension force, a point which is often overlooked, we derive the Young-Dupré equation which relates the contact angle to the three surface tensions at play on the contact line (the common frontier to three immiscible media), emphasizing that this relation should be interpreted as a continuity equation rather than a force balance equation. This reinterpretation eliminates many mistakes made on the identification of forces acting on the system.  
Moreover, we show that a safe strategy for the correct identification of acting forces consists in defining a control volume whose surface  is close to the contact line but does not contain it. We can then apply the force balance on this volume control, and then stretch its surface until it touches the contact line. 
%
Finally, we illustrate this approach in a variety of ways on a standard problem: the equilibrium height of capillary rise (Jurin's law).

\section{Surface tension: a force tangent to the interface}

\subsection{Macroscopic perspective}
We start our analysis by giving an experimental evidence that the restoring force associated with the creation of an elementary surface area between two immiscible fluids is tangent to it, without any knowledge on the microscopic origin of this force.
We
denote by $A$ the area of the interface, and consider a process whereby this
area undergoes a reversible change by an infinitesimal amount $dA$.
The energetic cost associated with this process corresponds to the work to bring molecules from bulk to the interface. Assuming that the surface concentration remains constant, this work is then directly proportional to $dA$,
 and so can be written as: $\delta W = \gamma \delta A$, which defines the surface tension $\gamma$ between the two fluids. A detailed thermodynamic treatment shows that the surface tension is defined as a surface contribution to the grand potential $\Omega$ \cite{Landau}, although the confusing term \textit{free energy} is used in most textbooks and scientific papers.
 
\begin{figure}[htb]
\begin{center}
\includegraphics[width=\linewidth]{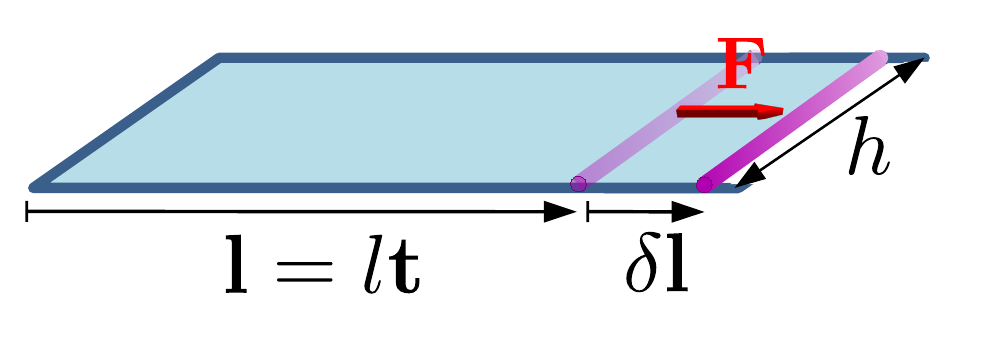}
\caption{Experimental evidence of the existence of a tensile force opposed to the creation of interface. The magnitude of the force is related to the surface tension $\gamma$ defined as the work to create a unit surface area.}
\label{tangential_force}
\end{center}
\end{figure}

\begin{figure}
\centering
\subfloat[]{\includegraphics[width=0.48\linewidth]{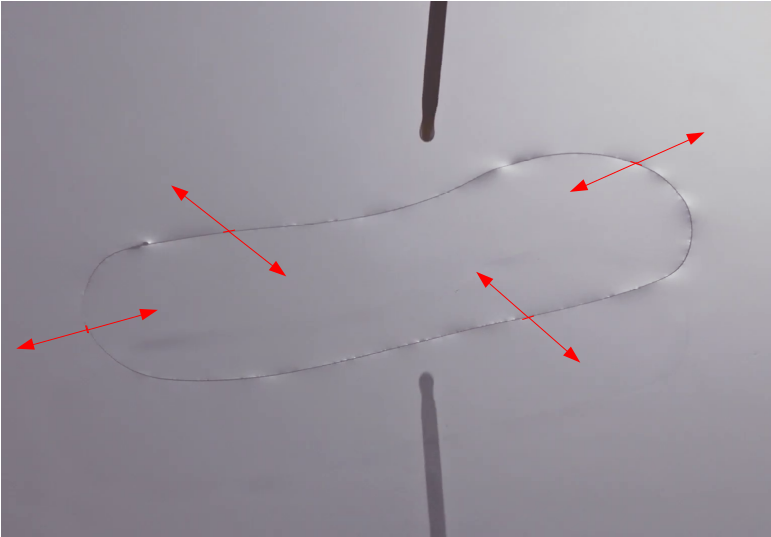}}
\hfill
\subfloat[]{\includegraphics[width=0.48\linewidth]{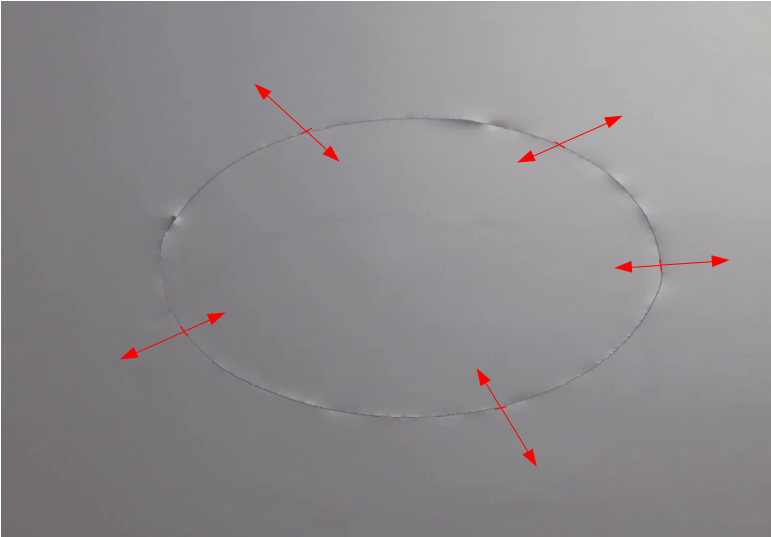}}
\caption{Experimental evidence of the tangential orientation of surface tension force: (a) a loop of thread is deposited on the surface of a water tank, with arbitrary shape. By symmetry, the forces (in red) acting on both sides of a thread element cancel each other. (b) After the addition of a drop of soap on the surface of the enclosed area, the thread forms a regular circular loop, because of the unbalance of inner and outer capillary forces. \label{fig:loop}}
\end{figure}

To illustrate the tangential nature of the restoring force, the experimental setup sketched in Fig.  \ref{tangential_force} is commonly invoked:
%
%
%
it consists of a U-shaped wire frame, on which is mounted a wire that can slide with
negligible friction. The frame and sliding wire support a thin film of liquid. Because
surface tension causes the liquid surface to contract, a force $\mathbf F$ is needed to keep the slider at fixed distance $l$ from the opposite edge. Then, the work associated with a small displacement $\mathbf{\delta l}=\delta l \mathbf t$ of the slider (where $\mathbf{t}$ is the unit vector tangent to the film) is $\delta W=\mathbf{F}\cdot\delta \mathbf{l}=F\delta l$.
This work is also equal to the increase of surface energy, $\delta W=2 \gamma \delta A$, with $ \delta A =  h \delta l $.  The factor $2$ reflects the two liquid-air interfaces of the film.
Identifying the two expressions of $\delta W$ yields
$\mathbf F = 2 \gamma h \mathbf{t}$. From the Newton's third law of motion, we conclude that the magnitude of the restoring force per unit length and per interface is $\gamma$. However, the demonstration with this experiment of the tangential orientation of the force acting on each interface is questionable: $\mathbf F$ is in fact the resultant of the forces acting on both liquid-air interfaces, which by symmetry must be tangent to the film, independently of the orientation of the force acting on each interface. A more convincing experimental evidence of the tangential nature of this force is the one shown in Fig. \ref{fig:loop}: a loop of thread is deposited on the surface of a water tank, with an arbitrary shape. Now delicately put a drop of soap on the surface circled by the thread. Immediately the loop adopts a regular circular shape, which results from the action of an isotropic force acting tangentially to the water surface on every point of the thread. Incidentally, this experience also highlights the fact that surface tension is an interfacial property that depends on the two fluids in contact: in Fig. \ref{fig:loop}(a), every elementary piece of thread experiences capillary forces from both the inner and outer surfaces, which have equal magnitude but opposite directions. In Fig. \ref{fig:loop}(b), the addition of soap decreases the magnitude of the inner tensile force, so that every piece of thread experiences a centrifugal force.

\subsection{microscopic perspective}
\begin{figure}[htb]
	\begin{center}
		\begin{overpic}[width=\linewidth]{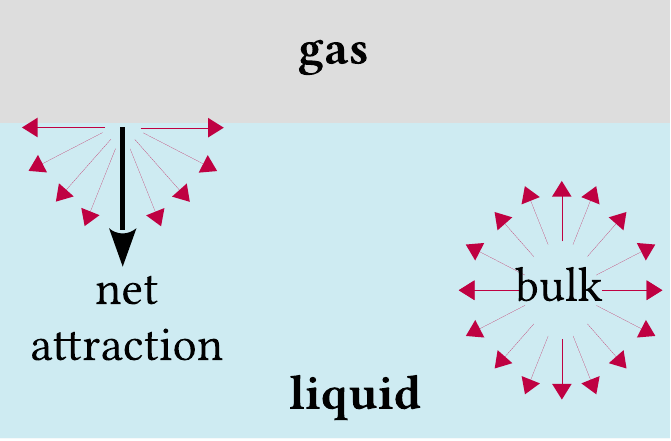}
			
		\end{overpic}
		\caption{Picture found in most textbooks to illustrate the origin of surface tension. This picture wrongly suggests that a molecule at the interface experiences a force perpendicular, and not tangential, to it.} 
		\label{wrong-picture}
	\end{center}
\end{figure}

The molecular origin of the phenomenon of surface tension evidently lies in the intermolecular cohesive forces. The usual explanation for the origin of surface tension from a microscopic perspective is a lack (for an interface between a liquid and its vapour) or a mismatch (for an interface between two immiscible liquids) of cohesive molecular bonds, as illustrated in Figure \ref{wrong-picture}, a picture which is found in most textbooks. This description yet suggests that a molecule at the interface experiences a net force oriented normally to the interface.  But then why is the surface tension force oriented tangentially to it ? A misleading justification is that the competition between all molecules pushing inwards results in a macroscopic tangential force. For instance, Adamson and Gast \cite{Adamson} uses the mechanical analogy of a weight acting on a rope and pulley to illustrate how the normal force is converted to a tangential tension. Yet, it is not clear what part plays the role of a pulley here.
Moreover, with such an argument the surface tension of a solid should be zero, since its constituents cannot freely move inwards like for a liquid. 

\begin{figure}
	\centering
	\includegraphics[width=\linewidth]{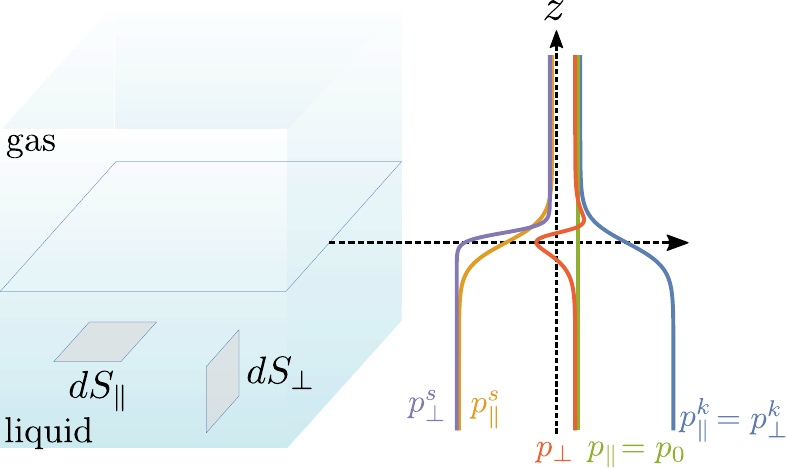}
	\caption{Microscopic origin of surface tension: the pressure force acting surface element $dS_\parallel$ parallel to the liquid-gas interface is constant, as the decrease of its kinetic contribution is exactly compensated by its static counterpart. The pressure force acting on the  surface element $dS_\perp$ perpendicular to the liquid-gas interface is negative in the vicinity of the interface, because the positive kinetic contribution decreases on a length scale around the interface comparable to the interatomic distance while the negative static contribution increases on a lengthscale comparable to the range of attractive interactions between molecules.}
	\label{fig:interface}
\end{figure}

Actually, the microscopic description above is incomplete, as it considers only the static part of the molecular interactions. The full microscopic interpretation of surface tension has been clearly exposed by M. V. Berry \cite{Berry}. We resume here his illuminating arguments. Let $p_\parallel(z)$ and $p_\perp(z)$ be the pressure acting on surface elements $dS_\parallel$ and $dS_\perp$, centered on a distance $z$ from the interface, and oriented parallel and normal to it, respectively (Fig. \ref{fig:interface}). By pressure, we mean here the normal component of stress acting on these surface elements.
Each of these pressures has two contributions: 
the first is the static contribution which arises because of the 
interactions between molecules. It is usually negative (so this contribution is indeed a tension), because the long-range attractive interactions between molecules take over their short-range repulsive interactions. This is the contribution which is mentioned in the partial microscopic description above.
The second is the kinetic contribution which comes from the transport of momentum by molecules moving
across the surface because of thermal agitation, and is always positive:
\begin{align}
p_\parallel(z)=p_\parallel^{s}(z)+p_\parallel^{k}(z), \quad \quad
p_\perp(z)=p_\perp^{s}(z)+p_\perp^{k}(z).
\end{align}
The kinetic pressure is the only pressure in an ideal gaz. As its value depends only on the local number density and temperature, it is isotropic: $p_\parallel^{k}(z)=p_\perp^{k}(z)$. However it is not homogeneous. If, for simplicity, one considers from now on an interface between a liquid and a gas; the kinetic pressure then continuously decreases when passing from the liquid to the gas phase.
The static part, on the other hand, is strongly anisotropic when approaching the liquid-gas interface, because of the long-range character of molecular interactions: for the surface element $dS_\parallel$, when approaching the interface, a molecule of the liquid phase interacts with fewer and fewer molecules located at the other side of the surface, because of the local density decrease. Thus, $p_\parallel^{s}(z)$ increases from its (negative) liquid bulk value to its (slighlty negative, or zero for an ideal gas) gas bulk value when crossing the liquid-gas interface, 
and the increase occurs over a distance comparable to the range of a few interatomic distances. Actually, the vertical balance of forces requires that the increase of $p_\parallel^{s}(z)$ exactly compensates the decrease of $p_\parallel^{k}(z)$ so that the total pressure $p_\parallel$ is a positive constant $p_0$, which is also the isotropic pressure far from the interface (we neglect here the pressure gradient caused by the gravity). Hence, $p_\parallel^{s}$, $p_\parallel^{k}$ and $p_\perp^{k}$ all three change significantly over a same spatial distance of a few intermolecular distances

For the element of surface $dS_\perp$, a molecule of the liquid phase still interacts with many molecules across the test surface when approaching the interface, 
and $p_\perp^{s}(z)$ remains close to its (negative) bulk value except in the extreme vicinity of the liquid-gas interface: the increase from the liquid bulk to gas bulk value occurs on a distance much shorter than for $p_\parallel^{s} (z)$. Therefore, there is a domain in the vicinity of the interface in which $p_\perp(z)<p_0$. The surface tension is defined as the difference between $p_\parallel(z)=p_0$ and $p_\perp(z)$ integrated over all $z$:
\begin{equation}
\gamma=\int_{-\infty}^{+\infty} \left(p_0-p_\perp(z)\right)dz.
\end{equation}
This equation simply states that a real interface can be replaced with an ideal interface where pressure is isotropic and uniform on both sides, plus a tensile force acting tangentially to the interface.

\section{Young-Dupré equation: an interface condition at the contact line}\label{sec:Young}
We now focus our attention on the Young-Dupré equation and its interpretation. When two immiscible fluids (e.g. a liquid and a gas) and a solid are brought in contact, the geometry of their connection is dictated by the Young-Dupré equation. The \textit{contact angle} is conventionally defined as the angle that the liquid-gas interface makes with the solid at the contact line (the locus of points at the frontier between the three media). The Young-Dupré equation relates the contact angle $\theta_c$ to the three surface tensions at play:
\begin{equation}
\gamma_{LG} \cos \theta_c +\gamma_{SL}-\gamma_{SG}=0.
\label{Young}
\end{equation}
\begin{figure}[htb]
	\centering
	\includegraphics[width=\linewidth]{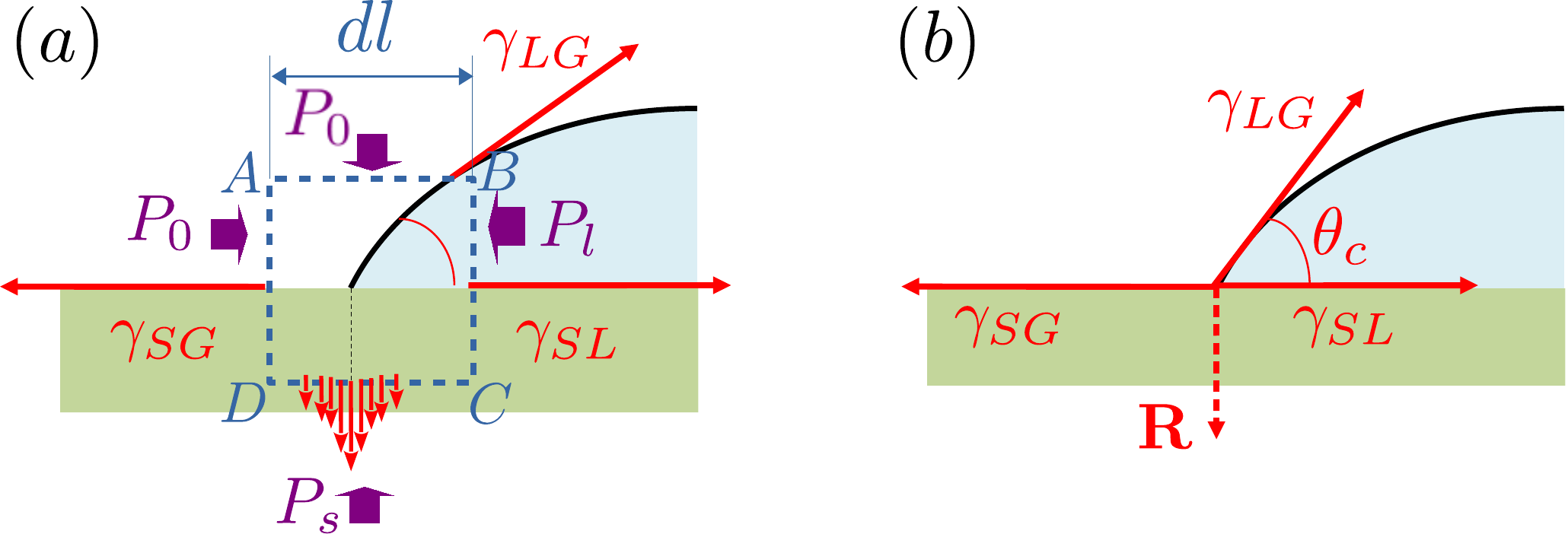}
	\caption{Young-Dupré equation derivation obtained from Newton's second principle : a) Forces acting on a vanishing control volume surrounding the contact line.
		b) Young-Dupré equation is an interface condition at the contact line: it relates the contact angle $\theta_c$ to the three surface tensions $\gamma_{SG}$, $\gamma_{LG}$, $\gamma_{SL}$. }
	\label{fig:Young}
\end{figure}

Young-Dupré equation is a source of much confusion for the students: first, it is inevitably introduced as the balance of forces acting on the contact line. This definition is indeed quite confusing: physical forces act on material systems, not on mathematical lines.
Moreover, in many textbooks the Young-Dupré equation is derived using arguments based on ``free energy'' minimization, and thus the support reaction $\mathbf{R}$ is even not mentioned or represented, yielding an unbalance of forces  in the vertical direction.

\begin{figure}[htb]
 	\centering
 	\includegraphics[width=.8\columnwidth]{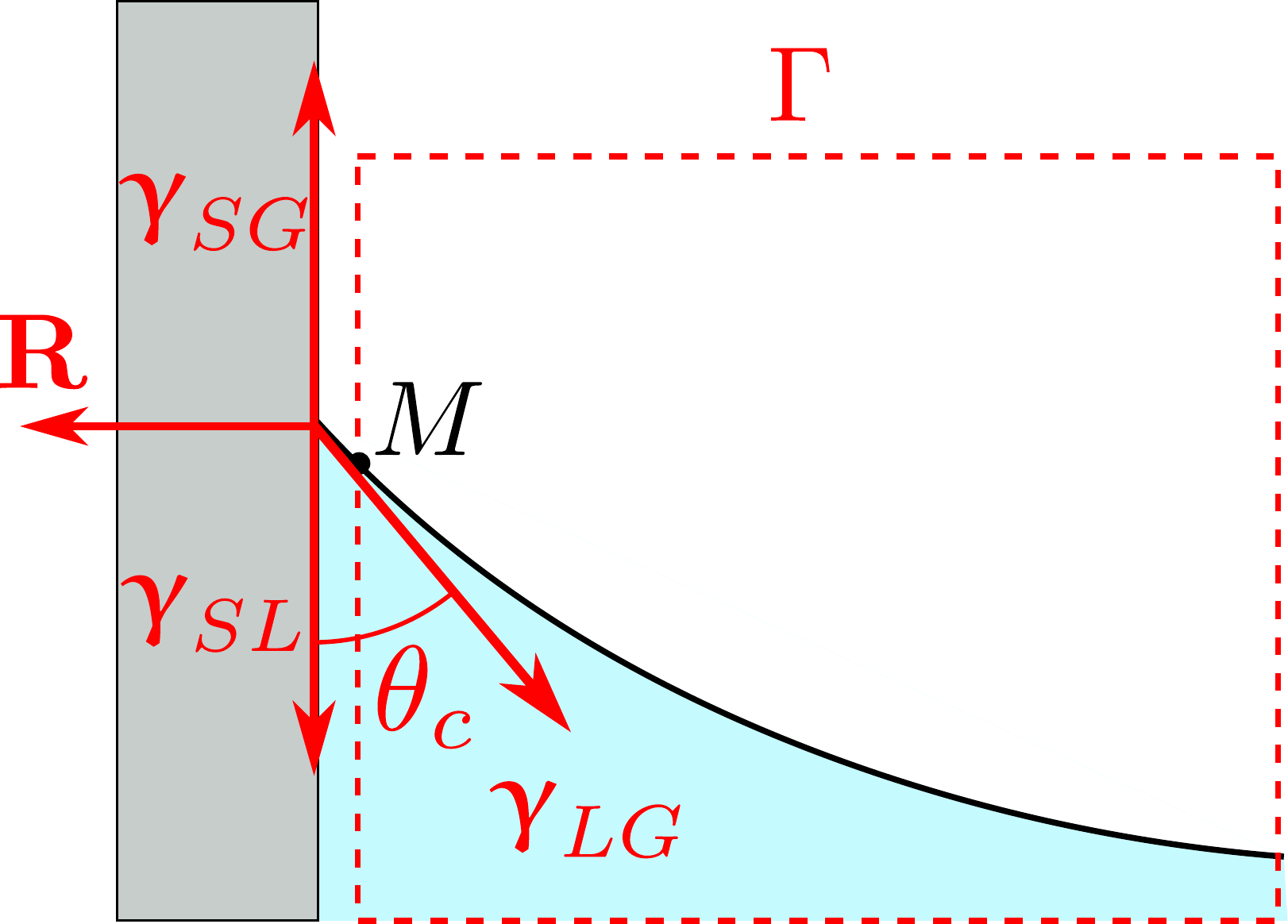}
 	\caption{Misconception of Young-Dupré equation: what is the force acting on the liquid-gas interface at the contact line ? \label{fig:misconception}}
 \end{figure}

Most importantly, the misinterpretation of Young-Dupré equation as a balance of forces blurs the identification of capillary forces acting on a system. To illustrate this point, consider the meniscus formed by a liquid at the vicinity of a vertical wall, as illustrated in Fig. \ref{fig:misconception}.
To the question \emph{what is the force exerted on the liquid-gas interface at the contact line ?}, the following (wrong) answers usually come out:\\ 
Answer \#1: since Young's law Eq. \ref{Young} precisely expresses the balance of forces at the contact line, and the contact line is precisely located at the end of the liquid-gas interface, the net force acting at the contact line at this position is $0$.\\
Answer \#2: the force acting on the liquid-gas interface at the contact line is caused by the wall. In virtue of the Newton's fundamental law of action-reaction, the exerted force must be $\mathbf R=-\gamma_{LG}\sin \theta_c \mathbf{n}$, where $\mathbf{n}$ is the unit vector normal to the wall. 

No one of these reasonings lead to the correct answer, which is $-\bm{\upgamma}_{LG}$. 
To understand why in Fig. \ref{fig:Young} four forces act on the contact line, which is at the end of the liquid-gas interface, while only one must be considered to answer correctly to the question, we derive in the following the Young-Dupré equation using a mechanical approach, and emphasizing that it should be interpreted as an interface condition rather than a balance of forces on a immaterial line. 
 This point of view will be helpful to unambiguously determine the capillary forces acting on a system.
Let us consider a  control volume with square section $dl\times dl$ enclosing the contact line, as shown in Fig. \ref{fig:Young}a. For simplicity, we suppose that the geometry is invariant by translation in the direction perpendicular to the figure.
The forces (per unit length) acting on this volume control can be divided in three categories:
\begin{itemize}
	\item forces acting on the volume, like the weight $\rho_{\text{eff}}dl^2\mathbf g$, where $\rho_{\text{eff}}$ is the effective density of the enclosed system. We  express them using the generic form $\rho_{\text{eff}} \mathbf f_v dl^2$. The key point is that they scale as $dl^2$.
	\item forces acting on the four faces: pressure is uniform in the gas and in the liquid, with respective values $P_0$ and $P_{l}$.
	The pressure within the solid has two contributions: the first one, noted $P_s$, equilibriates the pressure in the two fluids above and so its value continuously increases from $P_0$ in the region under the gas to $P_{l}$ under the liquid. 
	The associated pressure forces acting on the contour can be expressed using the generic form $\sum_i \mathbf{P}_i dl$, and scale as $dl^1$. The second contribution to the pressure within the solid is caused by the elastic response to the punctual normal force $\gamma_{LG}\sin \theta_c \mathbf n$ pulling on its surface. Usually surface tensions force are much weaker than the cohesive forces in the solid, and the deformations of the solid are imperceptible. Nevertheless, components of stress are finite: they spread through the solid material from the source point \cite{Landau_elasticity}. Conversely, when approaching the contact line, the stress conscentrates on a very localized zone, so its integration over segment $DC$ tends to a constant $\mathbf R$ independent of $dl$.
	\item capillary forces $\bm{\upgamma}_{LG}$, $\bm{\upgamma}_{SL}$, $\bm{\upgamma}_{SG}$ acting tangentially to the interfaces between the liquid, gas, and solid phases. These forces also scale as $dl^0$.
\end{itemize}
Application of Newton's second law yields:
\begin{align}
\sum_i \mathbf{P_i}dl+\rho_{\text{eff}}\mathbf f_v dl^2+\left( \bm{\upgamma}_{LG}+\bm{\upgamma}_{SL}+\bm{\upgamma}_{SG}+\boldsymbol{R} \right) \nonumber \\ =\rho_{\text{eff}} \mathbf{a}_{cm}dl^2
\end{align}
where the right-hand side term represents the inertia of the enclosed system; $\mathbf{a}_{cm}$ is the acceleration of its center of mass.
Taking this equation to the limit $dl \rightarrow 0$ finally yields:
\begin{equation}
\bm{\upgamma}_{LG}+\bm{\upgamma}_{SL}+\bm{\upgamma}_{SG}+\boldsymbol{R}=\mathbf{0}.
\label{balance}
\end{equation}
The vertical projection $R+\gamma_{LG}\sin \theta_c=0$ simply reflects the law of action-reaction (Newton's third law). 
The horizontal projection of this equation yields the Young-Dupré equation (Eq. \ref{Young}). This derivation is enlightening in two aspects:
first, it shows that rather a balance of forces on a immaterial line, the 
Young-Dupré equation is an interface condition, as there exist in other domains of physics (but here at the boundary between three medias), \eg such as the (disc)continuity of the tangential and normal components of the electric field at the interface between two media. As such, Young-Dupré equation dictates the geometry in the vicinity of the contact line: the contact angle is imposed by the values of the three surface tensions at play. 
Second, our derivation shows that in theory the Young-Dupré equation is not restricted to the static case; it applies to a moving contact line (with non uniform motion) as well. However, it is well documented that on real substrates the contact angle is not the same for an advancing or receding contact line, because of the surface imperfections \cite{de_Gennes}, a feature which is not captured by the Young-Dupré equation.

%


We now come back to the question of the force exerted on the liquid-gas interface at the contact line that we discussed at the beginning of this Section. 
First we can simply explain why the reasoning leading to answers \#1 and \#2 is wrong: in answer \#1 the insidious assumption was to consider the contact line as a material part of the liquid-gas interface, and that Young-Dupré equation expresses the balance of forces on it. As we stressed out, the contact line is an immaterial line, and thus does not experience or exert forces. The fallacy in answer \#2 is in considering that the system can be split in two independent subsystems: the solid wall in one hand, and the liquid and gas phases in the other hand. But this is wrong: the reunion of the two systems create new interfaces and so new capillary forces acting tangentially on them.

To answer to the initial question, let us consider a control volume $\Gamma$ which encloses the liquid and gas phases, except for a thin layer at the vicinity of the wall (see Fig. \ref{fig:misconception}). It is then clear that the capillary force exerted by this thin layer on the system at point $M$ tends to $-\bm{\upgamma}_{LG}$ as the layer thickness decreases to $0$.
More generally, to avoid any mistakes in the identification of forces acting on a system containing contact lines, 
we recommend to define first a control volume whose surface comes sufficiently close to, but does not embed the contact lines. We illustrate this approach in the next section.

\section{Application: Capillary rise (Jurin's law)}
We are now equipped to study capillary phenomena with the mechanical approach. In this section we use it to derive the expression for the equilibrium height of capillary rise, known as Jurin's law: when the lower end of a narrow tube is placed in a liquid such as water, a concave meniscus forms and the liquid climbs up to an equilibrium height $h$ which depends on the tube radius $r$ and the surface tension between liquid and air $\gamma_{LG}$.
The usual derivation based on thermodynamic approach is as follows: the change of surface ``free energy'' (grand potential, precisely) for an ascension at height $h$ corresponds by the replacement of a solid-gaz interface by a solid-liquid interface on that height: $F_{s}=\left( \gamma_{SL}-\gamma_{SGV} \right)2\pi r h$. According to Young-Dupré equation, it can rewritten as  $F_{s}=-\gamma_{LG} \cos \theta_c 2\pi r$. The change of gravitational energy is $F_{w}=\pi r^2 \int_O^h \rho g z dz=\pi r^2 \rho g h^2/2$. At Equilibrium, the total free energy $F_{tot}=F_{s}+F_{w} $ must be at a minimum: $d F_{tot}/dh=0$ \cite{note}. This yields the famous Jurin's law
\begin{equation}
h =\dfrac{2 \gamma_{LG} \cos \theta_c}{ \rho g r }.
\label{Jurin}
\end{equation}
Let us now turn to the mechanical approach: the equilibrium configuration is derived from the balance of forces acting on the system.
 To avoid misidentification of capillary forces acting on the system, we follow the recommendation set out in Section \ref{sec:Young}, and choose a control volume whose surface does not include the contact line.
 As a practice, we apply the force balance over various control volumes which satisfy this property, and show that they all lead to the expression (\ref{Jurin}).
 \begin{figure}[htb]
 	\begin{center}
 		\includegraphics[width=\linewidth]{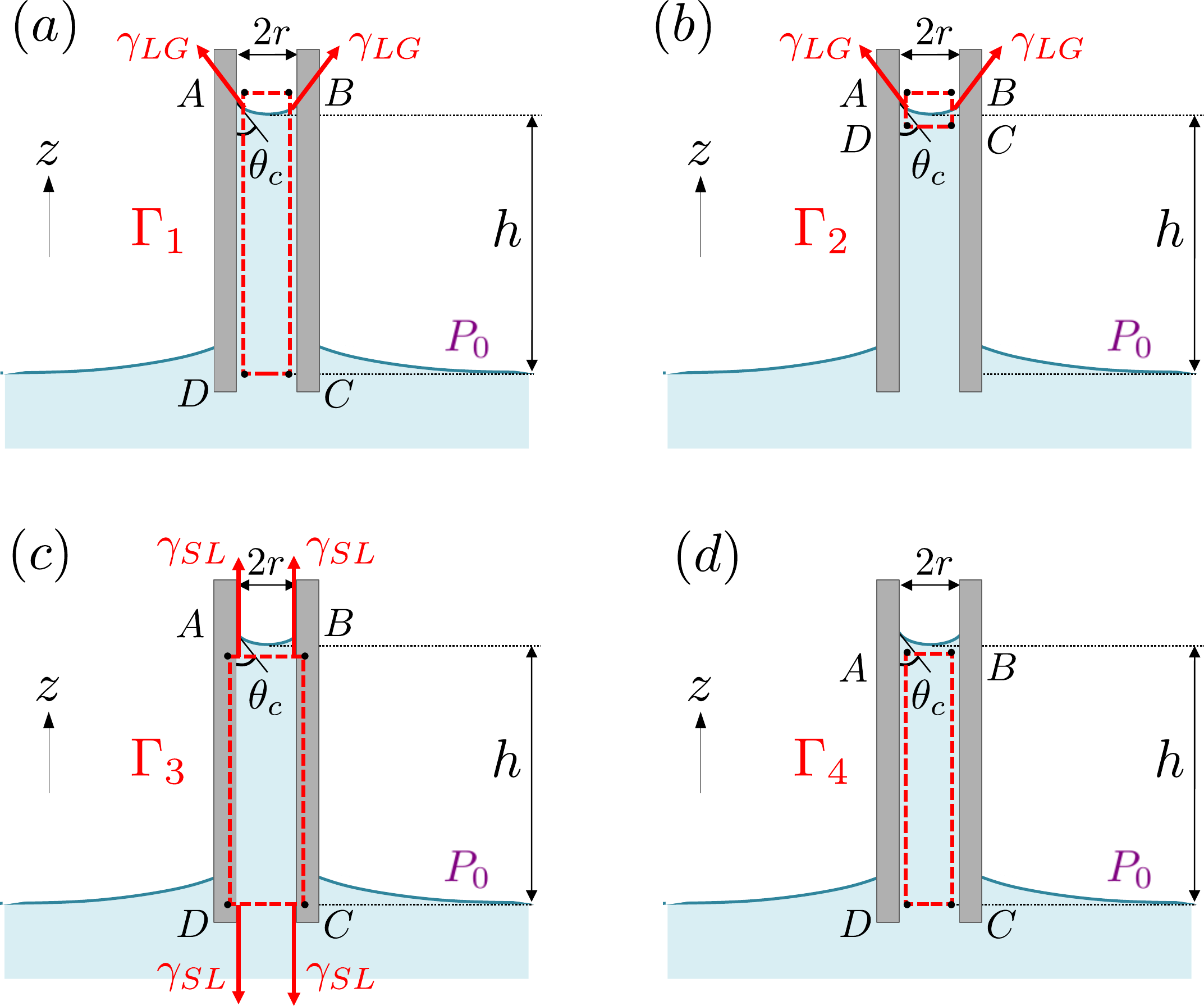}
 		\caption{Derivation of Jurin's law using mechanical approach (balance of forces) on four different control volumes.}
 		\label{fig:Jurin}
 	\end{center}
 \end{figure}
 \paragraph{Control volume $\Gamma_1$ (Fig. \ref{fig:Jurin}a) --}
 We first consider the axisymmetric control volume $\Gamma_1$ inside the capillary tube, whose cross-sectional area is delimited by the rectangle $ABCD$ (see Fig.  \ref{fig:Jurin}a). The segment $AB$ is located just above the meniscus (\ie at height $h+\epsilon$ with $\epsilon\rightarrow 0$), segments $BC$ and $AD$ inside the capillary tube, close from the walls, and segment $CD$ at the same height as liquid level in the pool far from the capillary tube. 
 As before, we identify the volume, surface and boundary (capillary) forces acting on this  control volume: the first one is the weight of the liquid column, $-\rho g h \pi r^2 \mathbf{e_z}$. The pressure forces acting on segments $AD$ and $BC$ cancel each other by symmetry. Similarly the pressure on segments $AB$ and $CD$ are both equal to the atmospheric pressure $P_0$ and so the associated forces cancel each other. The boundary forces are the capillary forces acting tangentially all along the meniscus contour, with constant norm $\gamma_{LG}$. Integration over that contour leads to the vertical force $2 \pi r \gamma_{LG}  \cos \theta_c \mathbf{e_z}$. Finally, balance of weight and capillary forces immediately lead to Jurin's law Eq.  \ref{Jurin}.
 
  \paragraph{Control volume $\Gamma_2$ (Fig. \ref{fig:Jurin}b) --}
  We now consider the control volume $\Gamma_2$, which is defined similarly to $\Gamma_1$ except that segment $CD$ is located just below the meniscus, \ie at height $h-\epsilon$ with $\epsilon\rightarrow 0$  \cite{note2}. The weight of the system is then negligible. Pressure forces on segments $AD$ and $BC$ still cancel each other (and are also negligible in the limit $\epsilon \rightarrow 0$). As before, the resulting capillary force is equal to $2 \pi r \gamma_{LG} \cos \theta_c \mathbf{e_z}$. It is balanced by the pressure forces acting on both sides of the interface, $(P_{CD}-P_{AB})\pi r^2  \mathbf{e_z}$ . Incidentally, introducing the radius of curvature  $R$ of the meniscus and the geometrical relation $R=r/\cos\theta_c$, one simply recovers Laplace's law. One has $P_{AB}=P_0$, while from hydrostatic law $P_{CD}=P_0-\rho g h$. Hence resulting pressure forces are $-\rho g h \pi r^2  \mathbf{e_z}$, and their balance with capillary forces finally leads to Jurin's law.
 
  \paragraph{Control volume $\Gamma_3$  (Fig. \ref{fig:Jurin}c) --}
  We then consider the volume control $\Gamma_3$ delimited by contour $ABCD$, where segment $AB$ is just below the meniscus, segments $BC$ and $AD$ lie within the walls of the capillary tube, in the vicinity of the solid-liquid interface, and segments $CD$ is positioned at the same height as the liquid level in the pool far from the capillary tube. As for the control volume $\Gamma_1$, the weight is $-\rho g h \pi r^2 \mathbf{e_z}$. Pressure forces on segments $AD$ and $BC$ still cancel each other by symmetry, but pressure on segment $AB$ and $CD$ are different: $P_{CD}=P_0$, and according to Laplace law, $P_{AB}=P_0-2\gamma/R$, where $R$ is the meniscus radius of curvature: $R=r/\cos\theta_c$. Capillary forces acting on segment $AB$ are located at the solid-liquid interface and are oriented tangentially to it. Integration over the contour gives $2\pi r \gamma_{SL}\mathbf{e_z}$. This resulting force is balanced by the exact same force with opposite direction acting on segment $CD$. Finally, Jurin's law results from the balance of weight and pressure forces.
  
  \paragraph{Control volume $\Gamma_4$ (Fig. \ref{fig:Jurin}d) --}
 Finally, we consider the volume control  $\Gamma_4$ which is identical to the previous one except that now segments $AD$ and $BC$ are in the liquid-side of the solid-liquid interface. Weight and pressure forces remain the same, while no capillary forces act on this control volume. Hence, as in the previous case, balance of weight and pressure forces yields Jurin's law.

\section{Conclusion}
To summarize, in this paper we discussed from a macroscopic and microscopic point of views the origin of the tangential orientation of the surface force. We then derived the Young-Dupré equation using the mechanical approach, and stressed out that it should be interpreted as a surface condition at the contact line, rather than a balance of forces: this equation dictates the geometry at the vicinity of the contact line by relating the contact angle to the three surface tension at play. This interpretation allows a proper identification of the capillary forces acting on a system, which is the primary mistake in the mechanical treatment of capillary phenomena. In particular, we stressed out that the studied system must not contain any contact line on its surface. Finally, applications of the mechanical approach using this methodology are given on a standard example: the capillary rise. We hope this work will be useful to the undergraduate students that find the mechanical approach more intuitive than the thermodynamic approach. Beyond that, this is one of the fascinating feature of physics that a same phenomenon can be described using alternative approaches. Understanding and mastering these different approaches is certainly a strength when facing new physics problems.

\textbf{Acknowledgments} I thank J. Derr and L. Viennot for many useful discussions, and the undergraduate physics students from the University of Paris whose many valuable questions have nourished this work.

\bibliography{biblio} 

\begin{mcitethebibliography}{11}
\bibitem[not()]{note}
${d^2F_{tot}}/{dh^2}=\pi r^2 \rho g>0$, ensuring that the extremum is indeed a
  minimum.\relax
\mciteBstWouldAddEndPunctfalse
\mciteSetBstMidEndSepPunct{\mcitedefaultmidpunct}
{}{\mcitedefaultseppunct}\relax
\EndOfBibitem
\bibitem[not()]{note2}
Segment $CD$ could be curved to closely follow the meniscus profile. This is
  not really necessary though, since $r$ is supposed to be much smaller than
  the capillary length $\sqrt{\gamma_{LG} / \rho g}$, so that the pressure is
  constant over the meniscus height.\relax
\mciteBstWouldAddEndPunctfalse
\mciteSetBstMidEndSepPunct{\mcitedefaultmidpunct}
{}{\mcitedefaultseppunct}\relax
\EndOfBibitem
\end{mcitethebibliography}
\bibliographystyle{rsc} 

\end{document}